\documentclass[onecolumn, draftcls]{IEEEtran}
\usepackage{bbm}
\usepackage{amsfonts}
\usepackage{tipa}
\usepackage{amssymb}
\usepackage{mathrsfs}

\usepackage{graphicx}
\usepackage{amsfonts,amssymb,amsmath}
\usepackage{latexsym}
\usepackage{hyperref}

\newtheorem{defn}{Definition}
\newtheorem{thm}{Theorem}
\newtheorem{lem}{Lemma}
\begin{document}
\baselineskip=22pt
\title{Efficient Signal-Time Coding Design and its Application in Wireless Gaussian Relay
Networks
\thanks{The corresponding author is Prof. Pingyi Fan}}
\author{\authorblockN{Pingyi~Fan\authorrefmark{1},~\IEEEmembership{Senior Member,~IEEE}, Pei~Sun \authorrefmark{1}, and Khaled~Ben~Letaief \authorrefmark{2},~\IEEEmembership{Fellow,~IEEE}}\\
\authorblockA{\authorrefmark{1}Department of Electronic
Engineering,\\
Tsinghua University, Beijing, P.R.China\\
E-mails : fpy@tsinghua.edu.cn}\\
\authorblockA{\authorrefmark{2}Department of Electronic and Computer Engineering,\\
Hong Kong University of Science and Technology, \\
Clear Water Bay, Hong Kong\\E-mail : eekhaled@ece.ust.hk}}
 \maketitle

\begin{abstract}\\
\baselineskip=18pt

Signal-time coding, which combines the traditional
encoding/modulation mode in the signal domain with  signal pulse
phase modulation in the time domain, was proposed to improve the
information flow rate in relay networks. In this paper, we mainly
focus on the efficient signal-time coding design. We first derive an
explicit iterative algorithm to estimate the maximum number of
available codes given the code length of signal-time coding, and
then present an iterative construction method of codebooks. It is
shown that compared with conventional computer search, the proposed
iterative construction method can reduce the complexity greatly.
Numerical results will also indicate that the new constructed
codebook is optimal in terms of coding rate. To minimize the buffer
size needed to store the codebook while keeping a relatively high
efficiency, we shall propose a combinatorial construction method. We
will then consider  applications in wireless Gaussian relay
networks. It will be shown that in the three node network model, the
mixed transmission by using two-hop and direct transmissions is not
always a good option.

\end{abstract}

\IEEEpeerreviewmaketitle

\baselineskip=24pt
\section{Introduction}
Relay network information transmission, as an important and
necessary technique, has been receiving significant attention in
recent years. However, only few results on the network capacities
and its efficient implementations were obtained for networks with
relatively simple topologies and special applications
\cite{yeung2000} \cite{cover91book}\cite{madsen05}. In the
literature, there are two different strategies to investigate relay
network information transmission. The first one is is based on
considering the networks with relatively simple topologies, such as
two-hop networks, three-node networks, diamond network, linear
networks and butterfly networks \cite{yeung2000} \cite{meulen71}
\cite{cover79}
 \cite{kramer-san}  \cite{fan-tcom09}. The second
one considers the networks composed of a lot of nodes and links
between them. It then employs random geometry and random graph
theory to find some statistical characteristics \cite{kumar}
\cite{fan-wang}. In this paper, we mainly investigate the three-node
networks.

It is well known that  the first Gaussian relay channel model was
introduced by \cite{meulen71}, and was thoroughly analyzed by Cover
and El Gamal in \cite{cover91book} \cite{cover79}  where they
presented upper and lower bounds for the general relay channel. In
\cite{KGG05}, the authors discussed different node cooperation
strategies and presented their corresponding capacity bounds. The
power allocation and capacity bounds for wireless relay networks
were discussed in \cite{madsen05}. On the capacity approaching
implementation, \cite{nabar} investigated the performance limits
using distributed space time coding in wireless relay networks.
Other works, e.g., \cite{liu}\cite{arnab05} \cite{duman} considered
the coding design and its implementation strategy for half-duplex
relay  channels. Network coding was also included in the discussion
for relay networks \cite{katti07}\cite{fan-tvt09}. Other discussions
on interference relay channels can be found in
 \cite{maric-isit08} \cite{jafar08it} .

 Recently, we considered the simple three-node network in a systematic way and found that it is
 possible to get a higher information flow rate beyond the known min-cut upper
 bound using conventional transmission techniques
 \cite{fan-tcom-sub09}. This observation led us to propose a new
 coding method, signal-time coding, which combines the traditional encoding/modulation
mode in the signal domain with the signal pulse phase modulation in
the time domain. Such a hybrid signal-time coding approach can be
considered as an integrated codec/modem processing in the two
dimensional combinatorial space: Signal domain and temporal domain.
The main feature of the proposed signal-time coding is that one can
separately design the codec/modem in the signal domain and  in the
time domain. Therefore, the well known efficient codec/modems, such
as LDPC, Turbo coding, and TCM  in the signal domain can be employed
. By doing so, the coding design problem becomes equivalent to that
of designing an efficient signal-time coding in the time domain. In
\cite{fan-tcom-sub09}, we provided an iterative construction method
of codebooks that can reach a derived lower bound of the coding rate
per pulse of signal-time coding. In this paper, we will further
investigate the efficient coding design. The main contributions are
as follows: 1) Explicit formulas are derived to find the maximum
available signal-time code number for an arbitrary code length,
which can then be used to reduce the code search time by exhaustive
computer search; 2) An iterative construction algorithm of
codebooks, which can reach the upper bounds in terms of coding
efficiency, is given; 3) A combinatorial codebook generation
strategy, which can keep a relatively high coding efficiency and
save the storing buffer size greatly, is developed; and 4) The
application of our proposed signal-time coding in wireless Gaussian
relay networks  is considered and investigated.

The rest of this paper is organized as follows. In Section II, we
shall revisit the three-node network model and introduce the
signal-time coding proposed  in \cite{fan-tcom-sub09}. We derive
explicit formulas to calculate the maximum available signal-time
codes in Section III and then present an iterative construction
algorithm of codebooks in Section IV. A combinatorial codebook
generation strategy and its coding efficiency are presented in
Section V. Some applications of signal-time coding in wireless
Gaussian relay networks are discussed in Section VI. Finally, we
present our conclusion in Section VII.

\section{System model and Signal-time coding}
Consider a three-node network as shown in Fig. 1.  The system
consists of three nodes, one source $S$, one relay $R$ and one
destination $D$.  For simplicity, assume that all the source
information must be forwarded by the relay node $R$ to its
destination $D$ and node $R$ can not transmit and receive signals
simultaneously. That is, the relay is  in half-duplex mode and there
does not exist a direct link between $R$ and $D$. Such kind of model
usually appears in wireline communications or satellite
communications, where the source node can not transmit information
to the destination node directly. In Section V, we shall discuss the
general case.

Let $C_1$ and $C_2$ denote the channel capacities of the links $S-R$
and $R-D$, respectively. For simplicity, we assume that the link
capacities of $S-R$ and $R-D$ are time invariant. In this case, one
wants to transmit some information from the source node  to the
destination node by relay forwarding. It is easy to see that the
maximum value of information flow $R_{max}$ satisfies
\begin{equation}
R_{max}\leq \min\{C_1,C_2\} \label{max-flow}
\end{equation}
by using the max-flow min-cut theorem.

Eqn. \eqref{max-flow} is the result obtained for a relay operating
in full duplex mode. It also seems to be  an upper of the achievable
information flow rate for relays in half duplex mode.

 Now let us consider the case of a relay working in
half duplex mode. For a given time period $T$, using the
conventional signal transmission method, the  maximum achievable
amount of information bits and its corresponding maximum average
information rate are given by the following lemmas.

\begin{lem}
Consider a simple relay network  with three nodes, the source $S$,
relay $R$ and the destination $D$. Also assume that the direct link
between the source and destination does not exist and that all the
information from the source to the destination need to be forwarded
by the relay. If the relay node $R$ is allowed to operate in half
duplex mode, then the maximum amount of information bits per period
of $T$ received by the destination is $\frac{C_1C_2}{C_1+C_2}T$ and
the maximum average value of information flow is
$\frac{C_1C_2}{C_1+C_2}$, where $C_1$ and $C_2$ represent the
channel capacities of the links $S-R$ and $R-D$, respectively.
\end{lem}

The outline of the proof of Lemma 1 can be given as follows. For
each time period $T$, a fraction of it is used to transmit signals
for the source to the relay, denoted as $T_1$. The remaining period
of time duration, denoted as $T_2$ is used for the relay to forward
its received signals to the destination. Suppose that all the links
can be employed to transmit information at their capacities,then we
have
\begin{eqnarray}
T_1+T_2&=&T \\
C_1T_1&=&C_2T_2
\end{eqnarray}

By solving the above two equations, we get
\begin{eqnarray}
C_1T_1=C_2T_2&=&\frac{C_1C_2}{C_1+C_2}T\\
T_1&=&\frac{C_2}{C_1+C_2}T \\
T_2&=&\frac{C_1}{C_1+C_2}T
\end{eqnarray}
This indicates that the maximum amount of information bits per
period of $T$ received by the destination is
$\frac{C_1C_2}{C_1+C_2}T$ and the maximum average value of
information flow is $\frac{C_1C_2}{C_1+C_2}$.

By using the results in Lemma 1, one can easily get Lemma 2.

\begin{lem}
Consider a simple relay network model with three nodes as described
in Lemma 1, then the maximum average value of information flow is
bounded by the following inequality
\begin{equation}
\frac{1}{2}\min\{C_1,C_2\}\leq \frac{C_1C_2}{C_1+C_2}\leq \min\{
\frac{1}{2}\max\{C_1,C_2\}, \min\{C_1,C_2\} \}. \label{twohop bound}
\end{equation}
\end{lem}

The results in Lemma 2 indicate that using conventional transmission
techniques, the achievable rate will not beyond the value of the
minimum one between $\frac{1}{2}\max\{C_1,C_2\}$ and
$\min\{C_1,C_2\}$. In the sequel, we shall refer to the right-hand
side of Eqn. \eqref{twohop bound} as two-hop upper bound.

By observing the results in Lemma 1, we find that in each period of
time $T$, only a part of the period with duration
$T_1=\frac{C_2}{C_1+C_2}T$ is used to transmit the signal from the
source $S$ to the relay $R$, and the remaining period of time
duration $T_2=T-T_1$ is used for the relay to forward the received
signals from the source $S$ to the destination $D$. One usually
would like to divide  each period of time $T$ into two subintervals.
The first one with length $T_1$ is employed for the source
transmitting signals to the relay, and the remaining part is used
for the relay to forward its received signals previously to the
destination. Obviously, this time division scheduling approach is an
efficient way to reach the so called maximum average value of
information flow for this simple relay network. However, it neglects
an important transmission characteristic of information flow. For
instance, consider the following scenario.

Suppose that $C_1=C_2$ and the period of time $T$ is divided into
four time slots. Based on Lemma 1, we know that the two slots are
used for transmitting signals from the source $S$ to the relay $R$,
and the other two slots for forwarding signals from $R$ to the
destination $D$. It is easy to see that there are four different
arrangements as shown in Fig. \ref{fig_2}, where the blank and
filled blocks represent the time slots for $S-R$ transmission and
$R-D$ transmission, respectively. According to the information flow
rule, $R$ will be permitted to transmit signals if it receives some
signals that it did not forward previously. We find that only cases
(3) and (4) in Fig. \ref{fig_2} are the available time scheduling
modes. Nevertheless, these two different available time scheduling
modes can bring 1 bit information and the corresponding information
rate with signal-time coding is at least $1/T$.

 The above example shows
that if we adjust the signal transmission time intervals for the
source to the relay and the relay to the destination while still
keeping the relay operating in half duplex mode, we will find that
there are various different scheduling modes that can be employed to
reach the so called maximum average value of information flow. In
addition, if we observe the information flow over the two hop
routing at the destination, it is not hard to see that the variation
of time scheduling modes is able to carry some information and such
time scheduling modes can be detected by the destination if each
time slot is long enough. Thus, the variation of time scheduling
modes can be employed to carry the information at the source node if
all the nodes have perfect knowledge of the whole network. As a
result, we can employ a hybrid way to encode and modulate the source
information. In other words, if the relay and destination employ
carrier sensing, then one can use rate splitting at the source node
by transmitting a part of the information using the time scheduling
mode and the remaining part using conventional signal transmission
way. By doing so, it is easy to see that the obtained average value
of information flow is given by the sum of the two different kinds
of information flows as shown in Fig. \ref{fig_3}, and is larger
than that given in Lemma 1.

The above  hybrid signal coding mode was called signal-time coding,
because it combines the traditional encoding/modulation mode in the
signal domain with the signal pulse phase modulation in the time
domain. One important point to be mentioned is that the main feature
of the signal-time coding is that one can separately design the
codec/modem in the signal domain and in the time domain. Therefore,
the well known efficient codec/modems, such as LDPC, Turbo coding,
and TCM in the signal domain can be employed here. The remaining
problem is that how to design an efficient coding mode in the time
domain.

Now let us consider the general case of $ C_1 \neq C_2$. In each
period time $T$, $T_1$ and $T_2$ are the total time durations of
employing links $S-R$ and $R-D$, respectively. By decomposing them
into $N$ time slots with different lengths, one can get
\begin{eqnarray}
T_1=\frac{C_2T}{C_1+C_2}=N\Delta t_1\\
T_2=\frac{C_1T}{C_1+C_2}=N\Delta t_2.
\end{eqnarray}
That is,
\begin{eqnarray}
\Delta t_1=\frac{C_2}{C_1+C_2}\frac{T}{N}\\
\Delta t_2=\frac{C_1T}{C_1+C_2}\frac{T}{N}.
\end{eqnarray}

Note that $C_1\Delta t_1=C_2\Delta t_2$. This implies that the
network transmits the same amount of information every active time
slot, which is independent of the links transmission status.

 For simplicity, we assume that both the relay and the
destination employ the same carrier sensing method so that they have
the same minimum resolution of signal detection in the time domain,
which we shall refer to $\Delta T$. Thus, we have $\min\{\Delta t_1,
\Delta t_2\}\geq \Delta T$.  After some manipulations, we get
\begin{equation}
\frac{C_1C_2}{C_1+C_2}\geq \max\{C_1,C_2\}\Delta T \frac{N}{T}
\label{delat T}
\end{equation}

Let $S_N$ denote the maximum number of available time slot order
sequences with length $2N$. Then, the information rate carried by
the signal-time coding in the time domain is given by $
f_{ct}=\frac{log_2(S_N)}{T}=A(N)\frac{N}{T}$, where
$log_2(S_N)=A(N)N$ and $A(N)$ is a weighting factor being dependent
on $N$.  By using Eqn. \eqref{delat T}, we have
\begin{equation}
f_{ct}\leq \frac{C_1C_2}{C_1+C_2}\frac{A(N)}{\max\{C_1,C_2\}\Delta
T} \label{timedomain rate}
\end{equation}
In fact, the upper bound in the right-hand side of Eqn.
\eqref{timedomain rate} can be reached when both the relay and
destination employ the minimum time resolution $\Delta T$.  As a
result, the total achievable information rate using signal-time
coding is given by
\begin{equation}
f_{ST}=\frac{C_1C_2}{C_1+C_2}(1+\frac{A(N)}{\max\{C_1,C_2\}\Delta
T})
\end{equation}
This  clearly shows that using signal-time coding can get a higher
information rate than that given in Lemma 1 and may be beyond the
two-hop bound given in Lemma 2 in some scenarios.

In this paper, we shall try to answer the following three questions:
1) What is the maximum available code number of signal time coding
for an arbitrary code length?; 2) How to construct an codebook with
relatively high efficiency and low memory units?; and 3) What is the
coding gain in different scenarios.

\section{Optimal Signal-time code design}
In Section II, we introduced signal-time coding. In this section, we
shall discuss the optimal signal-time coding design in terms of
coding efficiency. In fact, there exist two basic problems: 1) What
is the maximum number of the available signal-time codes with an
arbitrary length $2n$?; and 2) How to design it in a constructed
form ? To answer these two questions, we shall discuss them
separately in the following part.

\subsection{Maximum number estimation of signal-time codes}

\begin{figure}[h!]
\centering
\includegraphics[width = 10.0cm]{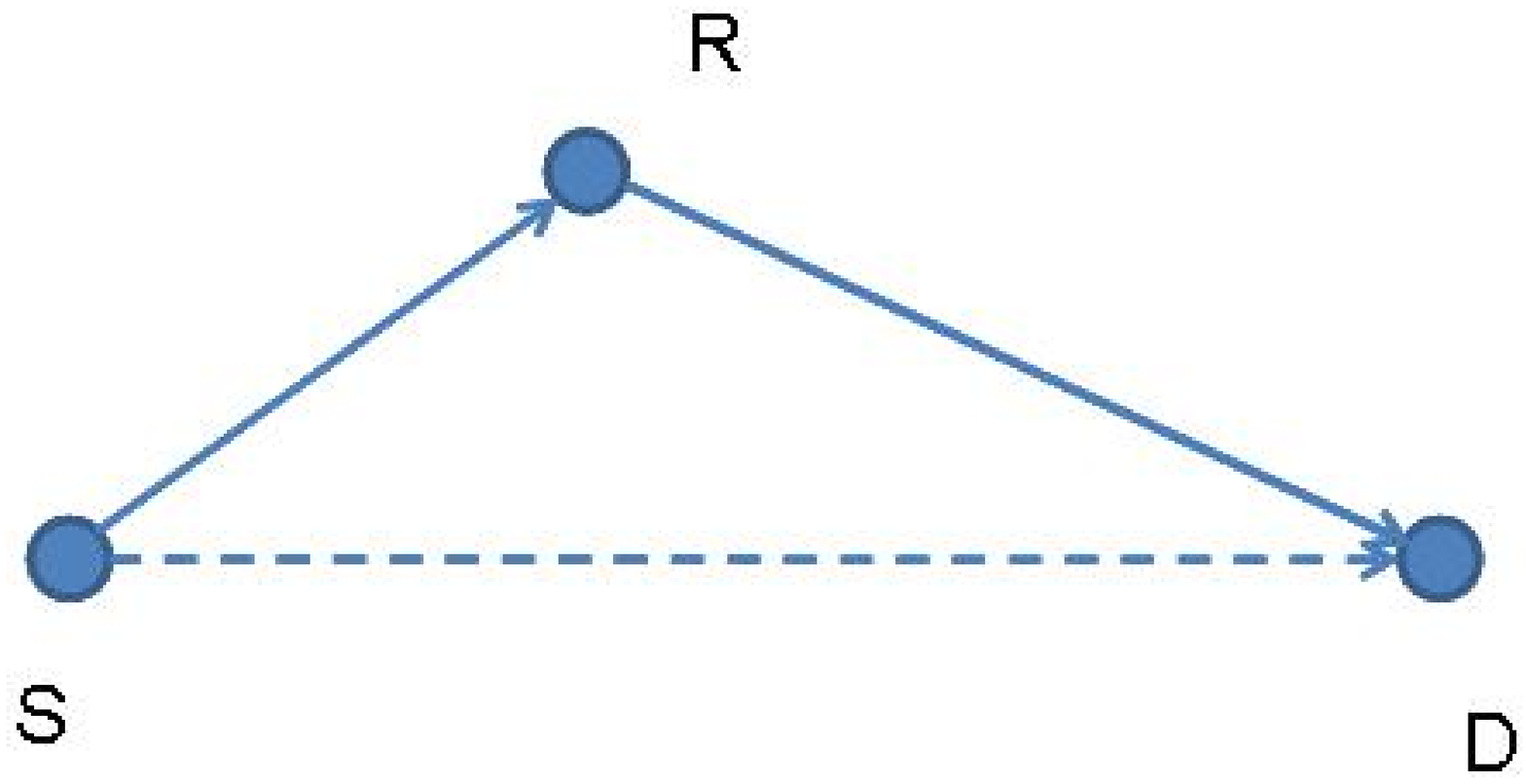}
\caption{Three node relay model }\label{fig_1}
\end{figure}

\begin{figure}[h!]
\centering
\includegraphics[width = 10.0cm]{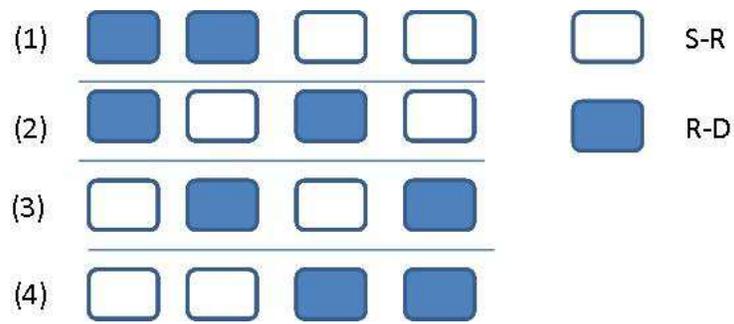}
\caption{Four time slots for signal time coding }\label{fig_2}
\end{figure}

\begin{figure}[h!]
\centering
\includegraphics[width = 10.0cm]{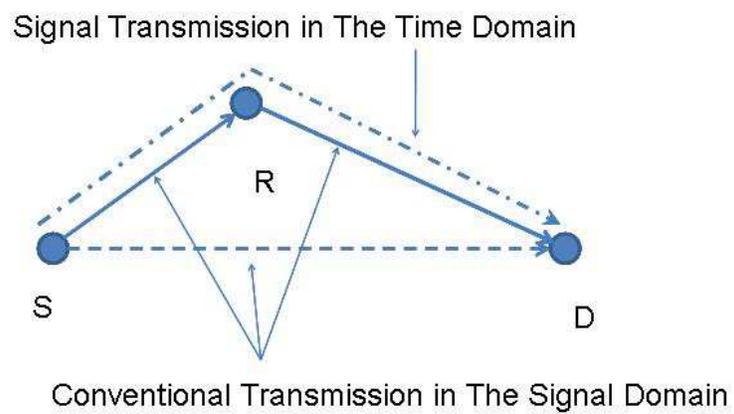}
\caption{Information transmission using signal time coding
}\label{fig_3}
\end{figure}

We begin by introducing a  lemma \cite{fan-tcom-sub09}.
\begin{lem}
Consider a relay network with three-nodes $S$,$R$ and $D$ as shown
in Fig. \ref{fig_1}, but without a direct link between $S$ and $D$.
Assume that the links $S-R$ and $R-D$ are with equal capacity $C$.
If $R$ is only permitted to operate in half duplex, then an
available time scheduling mode, which can achieve the average
information flow rate $C/2$, must satisfy the following condition:
For an arbitrary number $k$ ($0< k \leq 2n$), the appearance times
of $'1'$ is larger than or equal to that of the number $'0'$ in the
first $k$ time slots, where $'1'$ and $'0'$ represent the indicator
of a time slot being used by the links $S-R$ and $R-D$,
respectively.
\end{lem}

Note that this lemma can be directly obtained with the information
flow rule. That is, each available signal-time code must satisfy the
assignment rule given in  Lemma 1, where '1' and '0' represent two
kinds of time slots with different lengths if the link capacities of
$S-R$ and $R-D$ are different. Therefore, the code design can follow
the rule in Lemma 3. To do so, we need to introduce a definition.

\begin{defn}
Consider a '0' and '1' sequence with length $2n$. If it satisfies
the following two conditions
\begin{enumerate}
 \item  It has an equal number of '0' and
'1'. That is, the number of '0' is equal to that of '1' in the
sequence;
 \item  In any sub-sequence composed of the first $k$ ($k \leq 2n$) symbols, the number of '1' is not less than that of '0';
\end{enumerate}
 then, we we shall call the sequence as
 the available signal sequence.
\end{defn}

Based on this definition, the estimation of the maximum number of
available signal-time codes with length $2n$ can be obtained by
solving the following problem: How many available signal sequences
are there for a given length $2n$?

To answer this question, we shall define some symbols. let $S_n$
denote the total number of available signal sequences with length
$2n$, and let $F_n^k$ denote the total number of available signal
sequences with length $2n$, and whose first $k$ symbols are '1'.
Obviously, we have $ S_n=F_n^1$. In addition, by observing the first
two symbols in the available signal sequence, we get the following
\begin{equation}
S_n=S_{n-1}+F_n^2 \label{iterativeSn}
\end{equation}

Eqn. \eqref{iterativeSn} implies that the first symbol should be
assigned as '1'. If the second symbol is assigned as '0', the
following part can then follow for the available signal sequence
with length $2(n-1)$. Otherwise, the second symbol will be assigned
as '1'. In this case, the total number of available signal sequences
is $F_n^2$. Thus, the total number of the available signal sequences
is given in Eqn. \eqref{iterativeSn}.

Next we shall estimate $F_n^k$. By observing the property of
$F_n^k$, we have $F_n^k= 0$ if $ k>n$ and $ F_n^k= 1 $ when $ k=n$.
Consider the case when $k<n$. In this case, the available codes are
in the following form,

\begin{equation}
\underbrace{1\ \ 1\ \ \cdots \ \ 1}_k \underbrace{\ast \ \ \ast \ \
\cdots \ \ \ast}_{2n-1-k}0
\end{equation}
where $\ast$ denotes one symbol position, which may be filled by '0'
or '1'.

Let us decompose the marked part with $\ast$ into two different
subsections with the form
\begin{equation}
\underbrace{1\ \ 1\ \ \cdots \ \ 1}_k \underbrace{\square\ \ \square
\ \ \cdots \ \ \square}_k\underbrace{\ast \ \ \ast \ \ \cdots \ \
\ast}_{2n-1-2k}0 \label{de-form}
\end{equation}
where the $k$ symbols in the positions marked by $\square$ can be
selected as '0' or '1' randomly without any constraint, while the
selection of the remaining $2n-1-2k$ symbols marked by $\ast$ is
only dependent on the numbers of '0' and '1' in the positions marked
by $\square$, and is independent of their exact sequence order.

Without loss of generality, let $i$ ($0\leq i\leq k$) denote the
number of '1' in the positions marked by $\square$, then the number
of '0' in the positions marked by $\square$ is $k-i$, which implies
that the possible combination number is $C_k^i=\frac{k!}{i!(k-i)!}$.
In this case, the possible selection number of the remaining
$2n-1-2k$ symbols marked by $\ast$ is equal to the possible number
of the available sequence in the following form,

\begin{equation}
\underbrace{1\ \ 1\ \ \cdots \ \ 1}_{k+i} \underbrace{0\ \ 0\ \
\cdots \ \ 0}_{k-i}\underbrace{\ast \ \ \ast \ \ \cdots \ \
\ast}_{2n-1-2k}0
\end{equation}
By removing the first $k-i$ '0' and '1' pairs, we have
\begin{equation}
\underbrace{1\ \ 1\ \ \cdots \ \ 1}_{2i} \underbrace{\ast \ \ \ast \
\ \cdots \ \ \ast}_{2n-1-2k}0.
\end{equation}
Obviously, the above sequence length is $2n-1-2k+2i+1=2(n-k+i)$,
which implies that the possible number of the available sequence is
$F_{n-k+i}^{2i}$. By using the multiplier principle,  we get the
possible number of the available signal sequences in the form of
Eqn. \eqref{de-form} with $i$ '1' and $k-i$ '0' filled in the
$\square$ positions as $C_k^iF_{n-k+i}^{2i}$. By summing all the
cases with different $i$, we have
\begin{equation}
F_n^k=\sum_{i=0}^kC_k^iF_{n-k+i}^{2i} \label{Fkical}
\end{equation}
Note that in Eqn. \eqref{Fkical}, the equality $F_k^0=F_k^1=S_{k}$
is true for all positive integer $k$. In particular, when $k=1$, we
have
\begin{equation}
F_n^1=F_{n-1}^0+F_n^2=F_{n-1}^1+F_n^2.
\end{equation}
This is equivalent to Eqn. \eqref{iterativeSn}.

By using Eqns. \eqref{iterativeSn} and \eqref{Fkical}, we get
\begin{eqnarray}
S_n&=&S_{n-1}+F_n^2; \nonumber \\
F_n^2&=&C_2^0F_{n-2}^0+C_2^1F_{n-1}^2+C_2^2F_n^4, \nonumber \\
F_n^4&=&C_4^0F_{n-4}^0+C_4^1F_{n-3}^2+C_4^2F_{n-2}^4+C_4^3F_{n-1}^6+C_4^4F_n^8,\nonumber
\\
\cdots && \cdots \ \ \cdots \ \ \cdots \ \ \cdots \ \ \cdots \ \ \cdots\ \ \cdots \ \ \cdots \ \ \cdots \ \ \cdots \nonumber \\
F_n^{2^p}&=&\sum_{i=0}^kC_{2^p-1}^iF_{n-2^p+i}^{2i}+C_{2^p}^{2^p}F_n^{2^{p+1}}.
\label{iteration}
\end{eqnarray}
From Eqn. \eqref{iteration}, it is not hard to see that the
calculation of $S_n$ is dependent on $S_{n-1} ( or F_{n-1}^0) $ and
$F_n^2$, where $F_n^2$ is dependent on $F_{k}^{2i}, (1<k<n)$ and
$F_n^4$. Likewise, $F_n^4$ is dependent on $F_{k}^{2i}, (1<k<n)$ and
$F_n^8$. This iterative procedure can then follow. That is,
$F_n^{2^p}$ is dependent on $F_{k}^{2i}, (1<k<n)$ and
$F_n^{2^{p+1}}$. If $n \leq 2^q$ for some integer $q$, then the
iterative procedure will be stopped until $F_n^{2^q}$ due to
$F_n^{2^{q+1}}=0$. That is, $F_n^{2^q}$ is only dependent on
$F_{k}^{2i}, (1<k<n)$. As a result, we can use an iterative
algorithm to calculate $S_n$. The iterative procedure is given as
follows.
\begin{eqnarray}
Initialization &&\ \ \ F_1^0=1, \ \ F_2^0=1, \ \ F_2^2=2; \nonumber
\\
Using \ \  Eqn. \ \eqref{Fkical}&& iteratively \ \ to \ \ calculate
\ \ the  \ \ following \ \ terms
\nonumber \\
&&F_3^2 \longmapsto F_3^0 \ \ \ (F_3^{2k}=0, k>1) \nonumber \\
&& F_4^4 \longmapsto F_4^2 \longmapsto F_4^0 \ \ \ (F_4^{2k}=0, k>1)
\nonumber \\
&& \cdots \ \ \cdots \ \ \cdots \ \ \cdots \ \ \cdots \ \ \cdots \ \
\cdots\ \ \cdots \ \ \cdots \ \ \cdots \nonumber \\
&& F_n^{2^p} \longmapsto F_n^{2^{p-1}}\longmapsto \cdots \ \ \cdots
\longmapsto F_n^2 \longmapsto F_n^0 \ \ (2^p \geq n, \ \ F_n^{2k}=0,
k \geq 2^{p+1})  \nonumber
\end{eqnarray}

It is easy to see that using the above algorithm, we can save much
time to estimate the maximum number  of the available codes for
signal-time coding  compared to using exhaustive computer search.

\subsection{Codebook construction}

Consider the derivation of Eqns. \eqref{iterativeSn} \eqref{de-form}
and \eqref{Fkical}, we find that for a given code-length $2n$, the
construction of codebook with the highest coding efficiency can be
iteratively built up by using the codebooks with relatively shorter
lengths.  Table 1 shows the constructing process, in which $CBF_n^k$
represents the sub-codebook with length code-length $n$ and code
number $F_n^k$. Obviously, any pair of codes in the constructed
codebook with code-length $2n$ are different. This means that the
constructed codebook are uniquely decodable. We shall state this
more formally in a theorem.

\begin{thm}
For any positive integer $n$, the constructed code book with code
length $2n$ using the method as shown in Table \ref{table} is
uniquely decodable.
\end{thm}

This theorem indicates that the codebook constructed as shown in
Table \ref{table} can be used to perform signal-time coding.
Therefore, the optimal information bit numbers per pulse with
different code-length are presented in Fig. 4. It is known that a
loose upper bound of the coding rate on the signal-time coding is 1
bit/per pulse. The numerical results are very close to the upper
bound when the code lengths are large enough. i.e. $2n=300$, the bit
rate per pulse is 0.9611. However, due to the limited capability of
the computer search, we are not able to provide what is the exact
upper bound of the coding rate per pulse for signal-time coding.
Fortunately, by various numerical tests, we found that
$S_n=\frac{C_{2n}^n}{n+1}$ is true for all $n<150$!  If this is true
for all $n$, then using the Stirling formula,
\begin{equation}
n! \approx  n^{n+1/2} e^{-n}\sqrt{2\pi},
\end{equation}
we can get
\begin{equation}
\frac{C_{2n}^{n}}{n+1} \approx 2^{2n} \frac{1}{(n+1)\sqrt{\pi n}}
\end{equation}
and
\begin{equation}
\lim_{n\longrightarrow\infty}\frac{log_2\frac{C_{2n}^n}{n+1}}{2n}=1.
\end{equation}
As a result, we can conjecture that 1 bit/per pulse is a tight upper
bound of the coding rate per pulse for signal time coding. However,
this still needs to be proved.

\section{Efficient signal-time coding design}
In Section III, we presented the optimal code design. However, as
the code length increases, the codebook storage will become a
serious problem because the number of codes will increase
exponentially with respect to the code-length.  In this section, we
shall propose a combinatorial method, which combine a set of
codebooks with relatively shorter code-lengths to compose a codebook
with an arbitrarily required code-length. This combinatorial method
can be considered as an universal combinatorial method.

The construction steps are described as follows.
\begin{enumerate}
 \item  Select a relatively large positive integer $L$;
\item Construct all the codebooks with code length not greater than
$L$ by using the iterative method in Section III, and let $
CBST_{k}$ denote the codebook with code-length $k$, ($ 1\leq k\leq
L$);
\item Construct a codebook with code-length greater than $L$.

The
basic idea is that we shall construct  the codebook with code-length
$2n$ by concatenating some component codes. Here we give a simple
construction method.

Suppose that the required code-length is $2n$, then we get
$2n=mL+r$, where $m$ and $r$ are two positive even integers and
$0\leq r <L$. Based on this result, the codebook with code-length
$L$ and that with code-length $r$  obtained in Step 2) are selected
as the two component codes. Each code with code-length $2n$ is setup
in the following form.

\begin{equation}
\underbrace{C_1(L),C_2(L),\cdots, C_m(L)}C_{m+1}(r)
\label{component}
\end{equation}
where $C_k(L)$ and $C_k(r)$ denote the selected codes with length
$L$ and $r$ arranged in a fixed order of $k$, respectively. By
collecting all the codes in the form Eqn. \eqref{component}, we can
obtain the codebook with code-length $2n$.
\end{enumerate}

Using the concatenated coding, it is easy to evaluate the average
coding rates and its storage sizes for an arbitrarily length
codebook. The average coding rate per pulse $R_c(n)$ is given by
\begin{equation}
R_c(n)=\frac{mLR_L}{2n}+\frac{rR_r}{2n}
\end{equation}
where $R_L$ and $R_r$ denote the coding rates per pulse for the
optimal codebooks obtained in Section III with length $L$ and that
with length $r$, respectively.

In order to clearly characterize the coding rate and its buffer
size, we introduce two measures.

\begin{defn}
The coding rate loss coefficient of one codebook with code-length
$2n$ is defined by
\begin{equation}
\rho_R=1-\frac{R(n)}{R_n}
\end{equation}
where $R(n)$ and $R_n$ represent the coding rate of one codebook
with code-length $2n$ and that of the optimal coodbook with the same
code-length obtained in Section III.
\end{defn}

\begin{defn}
The code storage ratio of one codebook with code-length $2n$ is
defined by
\begin{equation}
\rho_M = \frac{S(n)}{S_n}
\end{equation}
where $S(n)$ and $S_n$ represent the required storing code number of
one codebook with code-length $2n$ and that of the optimal coodbook
with the same code-length obtained in Section III.
\end{defn}

Figs. 4-6 present some sample results on the coding rate, coding
rate loss coefficients and code storage ratios with different values
of $L$. From the results of Figs. 5 and 6, we get the following
observations: 1) As $L$ increases, the code rate loss ratios will
decrease. In particular, when $L$ is 60, the code rate loss ratio is
less than 6 percent for the code-lengths in the interval of 2 to
300; and 2) The combinatorial codebook storage  will reduce greatly
compared to that of the optimal codebook with the same length. For
example, when $L=60$ and $n=100$, the combinatorial codebooks
storing only requires about $1/10^{23}$ times memory units of the
optimal codebooks with the same length while the corresponding
coding rate loss is about 4 percent. This indicates that the
universal combinatorial coding method will bring a good tradeoff
between the coding rate loss and the codebook storing.

\section{Application in wireless Gaussian relay networks}

In Sections III and IV, we mainly discussed the signal-time coding
design under the assumption that the relay network topology adopts a
two-hop model. In this section, we shall consider the general
three-node model as shown in Fig. 3. The main difference from the
two-hop three-node model is that there exist a direct link between
$R$ and $D$. By using the general model, we shall discuss the
benefit of signal-time coding in wireless Gaussian relay networks.

In \cite{madsen05}, an upper bound on the average information flow
rate in time division relaying was given. Later, the achievable rate
for the general half-duplex Gaussian relay channel with the
decode-and forward protocol at the relay $R$  were presented in
\cite{aazhang07}. To observe the coding gain of signal-time coding
over the conventional signal transmission technique, we employed the
achievable rate for the general  half-duplex Gaussian relay channel
with the decode-and forward protocol at relay $R$ given in
\cite{aazhang07} as the baseline. The achievable rate for the
general half-duplex Gaussian relay channel with the decode-and
forward protocol is introduced here as a lemma.

\begin{lem}
For the general half-duplex Gaussian relay channel, the decode-and
forward protocol at relay $R$ achieves the following rate,
\begin{eqnarray}
R_{GC}&=&\sup_{0\leq t, r\leq 1}\min\{
tC(P_{SR})+(1-t)C((1-r^2)P_{SD_2}),\nonumber \\
&&tC(P_{SD_1})+(1-t)C(P_{SD_2}+P_{RD}+2r\sqrt{P_{SD_1}P_{RD}})\}
\label{HFAchievable}
\end{eqnarray}
where the power spectral density of Gaussian additive noises at the
relay and destination are both normalized as 1, and $P_{SR}$ and
$P_{RD}$ denote the received powers at relay $R$ from link $S-R$ and
at destination $D$ from link $R-D$, respectively; $P_{SD_1}$ and
$P_{SD_2}$ represent the received power at destination $D$ from link
$S-D$ directly in the time intervals of $R$ in the receiving and
transmitting modes, respectively; $r$ is the correlation between the
source and relay signals in multiple access mode and $C(x) =
\frac{1}{2} \log(1 + x)$ is the capacity of a Gaussian link.
\end{lem}
Note that the results in Lemma 4  are based on the use of the
normalized bandwidth of the transmission. In the following
comparison, we will consider the effect of transmission bandwidth.

Assume that we  employ the following transmission strategy: 1) At
the beginning of each transmission period $T$, the information at
source $R$ will be decomposed into three parts by rate splitting.
The first part will be transmitted by the direct link of $R-D$ using
the conventional signal transmission technique. The second part will
be transmitted over links $S-R$ and $R-D$ in a time division mode,
and the third part will be transmitted by signal-time coding in the
time domain. 2) The time division mode is adopted here. We first
divide the time period $T$ into $2n$ time sub-slots. In each
receiving sub-slot of relay $R$, the source transmit two different
information steams to $R$ and $D$ over links $S-R$ and $S-D$,
respectively, with different transmission powers, and in each
transmitting sub-slot of relay $R$, the source $S$ will stop its
signal transmission, then the destination $D$ only receives the
forwarded information from $R$, while it always detects the signal
pulses in the time domain and receives the carried information by
signal-time coding in the time domain regardless of the time
sub-slot being used by relay for receiving or transmitting.  In this
way, we use  three information sub-flows to transmit signals. The
total information amount in time period $T$ is given by

\begin{equation}
I_{A}=I_{A1}+I_{A2}+I_{A3}
\end{equation}
where
\begin{eqnarray}
I_{A1}&=&\frac{C_1C_2}{C_1+C_2}T  \\
I_{A2}&=&\frac{C_1C_2}{C_1+C_2} \frac{1.9222T}{\max \{C_1,C_2 \} \Delta T}  \\
I_{A3}&=&\frac{C_2C_3}{C_1+C_2}T
\end{eqnarray}
and $C_1$, $C_2$ and $C_3$ represent the link capacities of $S-R$,
$R-D$ and $S-D$, respectively; and $\Delta T$ is the minimum
required time resolution for signal detections at relay $R$ and
destination $D$. In fact, $I_{A1}$ is the information amount
transmitted over a two-hop routing of $S-R$ and $ R-D$ in the time
period $T$, $I_{A2}$ is the information amount transmitted by
signal-time coding in the time domain in the duration of $T$ where
we use the obtained result of the bit rate per pulse $0.9611$ is
employed and $I_{A3}$ is the information amount transmitted over the
direct link $S-D$ in the time period $T$.

The corresponding average achievable information rate by using the
proposed signal-time coding is given by
\begin{equation}
R_{ST}=\frac{C_1C_2}{C_1+C_2}(1+\frac{1.9222}{\max \{C_1,C_2 \}
\Delta T})+\frac{C_2C_3}{C_1+C_2} \label{ST-Infrate}
\end{equation}

In order to clearly characterize the coding gain of signal-time
coding over Gaussian relay channels, we make the following
assumptions.

1) In each time slot, the total transmission powers are limited by a
constant $P$.

2) The distance between $R-D$ is denoted by $d$ while the sum of the
distances from $S$ to $R$ and that from $R$ to $D$ is equal to $ad$
where $a>1$. Obviously, the relay is at a point of an ellipse with
$S$ and $D$ as its two focuses. This will help us observe the effect
of the distance variation between $S$ to $R$ (or the ratio between
the other two sides of the triangle) on the information flow rate.

3) The propagation path loss exponential factor $\alpha$ is set up
as 2.

4) The source and relay employ the same transmission frequency band,
and the bandwidth is denoted as $B$.

5) The normalized power of the additive Gaussian white noises (AWGN)
are set as 1. That is, $BN_0=1$ where $N_0$ is the power spectral
density of AWGN.

Based on the above assumptions, the parameters in Eqn.
\eqref{HFAchievable} are given by
\begin{eqnarray}
P_{SR}=\frac{P}{d_1^{\alpha}}, & &
P_{SD_1}=\frac{P}{d_3^{\alpha}}; \\
P_{SD_2}=\frac{(1-\beta)P}{d_3^{\alpha}},& & P_{RD_2}=\frac{\beta
P}{d_2^{\alpha}}.
\end{eqnarray}
while the parameters in Eqns. \eqref{ST-Infrate}, and \eqref{twohop
bound} are given by, respectively,
\begin{eqnarray}
C_1&=&Blog_2(1+\frac{\zeta P}{d_1^{\alpha}}), \\
C_2&=&Blog_2(1+\frac{P}{d_2^{\alpha}}), \\
C_3&=&Blog_2(1+\frac{(1-\zeta) P}{d_3^{\alpha}}),
\end{eqnarray}
and
\begin{eqnarray}
C_1&=&Blog_2(1+\frac{P}{d_1^{\alpha}}) \\
C_2&=&Blog_2(1+\frac{P}{d_2^{\alpha}})
\end{eqnarray}
 where $B$ is the transmission bandwidth, $d_3 =ad$, $d_1= \kappa d_3$,
$d_2=(1-\kappa)d_3$, and $0\leq \beta,\zeta \leq 1$.

In this case, the maximum achievable information rate using
signal-time coding for a given normalized time resolution $B\Delta
T$ is given by
\begin{equation}
R^{opt}_{ST}(B\Delta T)=\sup_{0\leq \zeta \leq
1}\{R_{ST}(\zeta,B\Delta T)\}
\end{equation}
and the two hop upper bound is given by
\begin{equation}
U_{two}=\min\{U_1, U_2 \}
\end{equation}
where
\begin{eqnarray}
U_1&=&\min\{Blog_2(1+\frac{P}{d_1^{\alpha}}),
Blog_2(1+\frac{P}{d_2^{\alpha}})\} \\
U_2&=&\frac{1}{2}\max\{ Blog_2(1+\frac{P}{d_1^{\alpha}}),
Blog_2(1+\frac{P}{d_2^{\alpha}}).
\end{eqnarray}

To guarantee fairness, $R^{opt}_{ST}(B\Delta T) $ and $U_{two}$ are
required to be normalized by $2B$. For simplicity, we denote them as
$R^{opt}_{N-ST}(B\Delta T)$ and $U_{N-two}$, respectively. Here we
also need to introduce a new concept, the coding gain of signal-time
coding, which is defined as the ratio of the maximum achievable
information rate using signal-time coding to that
 using the conventional transmission technique. It is given by
\begin{equation}
\gamma = \frac{R^{opt}_{N-ST}}{R_{GC}}
\end{equation}

Various numerical investigations have shown that using signal time
coding will bring more benefits in terms of information flow rate
compared with that only using the conventional transmission
techniques in the signal domain. That is, the coding gain is always
larger than 1. Due to space limitation, we do not include the result
of these investigations here. One can observe some of these from the
results in Figs. 7-10.

We shall now use some examples to illustrate the increment of
information flow rate by using signal time coding. Let $d=10$,
$\kappa=0.35$ and $0.75$, and $a=1.5$ and $2$. The value range of
$P$ is selected from $1$ to $50$ dB, which indicates the received
SNR variation over the direct link S-D is from $-19$ to $30$ dB. Two
normalized time resolutions are considered here, which are selected
as 6 and 12. Figs. 7-10 show some comparison results with time
division achievable upper bound and the two hop upper bound, where
the SNR used is that over the direct link $S-D$. Fig. 11 shows the
optimal power allocation of signal-time coding over different links.
From these results, we get the following observations.

1) When the normalized time resolution is relatively small, the
signal-time coding will provide more information rate gains. Thus,
reducing the required normalized time resolution will get more
benefits in terms of information flow rate.

2) When SNR is relatively low and the relay is relatively close to
the source, the relative coding gain of signal-time coding is bigger
and as the SNR increases, the relative coding gain will become
smaller. This suggests that when SNR is relatively small, we need to
adopt signal time coding to get more coding gain in terms of
information flow rate. Comparing the results in Figs. 7 and 9, we
can see that if the sum distances of $d_1$ and $d_2$ is relatively
small, the relative coding gain of signal-time coding decreases more
slowly as SNR increases.

3) As the relay is relatively far away from the source, if SNR is
relatively low, then using signal time coding will get a smaller
information rate gain compared to that case when the relay is
relatively close to the source. In contrast, when SNR is high
enough, using signal time coding will get more information rate
gains. And the gain will become bigger along with the increase of
SNR. This suggests that we need to employ signal-time coding when
the relay is relatively far away from the source. A more important
point is that the valid range of SNR using signal-time coding is
from a few dB to infinite when $\kappa=0.75$.

All the above observations can be explained from the results in Fig.
11.  When relay $R$ is relatively close to the source, it needs to
adjust its transmission power allocation to the different links
$S-R$ and $S-D$ in the receiving time-slots of relay $R$ to get its
maximum information rate. Fig. 11 indicates that when $a=2$, about
29 percent power will be allocated to the link $S-R$, so that its
capacity can have a good match with that one over link $R-D$, and
other 71 percent will be allocated to the $S-D$ link when the
normalized time resolution, NRT, is equal to 6 and SNRs are from -19
to -1 dB. As the normalized time resolution, NRT, increases, the
corresponding valid SNR range will expand. Due to space limitation,
we do not show this result. As the SNR is relatively high, the
optimal power allocation strategy is to allocate all the power to
the direct link $S-D$ and not to the link $S-R$. This implies that
the transmission system will not use the relay again to transmit
information.

In contrast, when the relay $R$ is relatively far from the source,
the analysis will become more complicated. It is  dependent not only
on the sum of the distances of $S-R$ and $R-D$, but also on their
relative ratio $\kappa$.

Let us observe the case $a=2$ first. In this case, it needs to
allocate the transmission power to different links $S-R$ and $S-D$
in the receiving time slots of relay $R$. Fig. 11 indicates that
when SNR is very low, all the power will be allocated to link $S-R$
and not to the direct link of $S-D$. This implies that the
transmission system would like to select  two hop transmission,
which further confirms that our considered simple three-node model
with direct link of $S-D$ in Section II is reasonable in some
scenarios even it may have a direct link between the source and
destination. Fig. 11 also indicates that when SNR is relatively low,
it will have a balance point on the power allocation to the link
$S-R$ and $S-D$, (see SNR = -11 dB, NRT=6, and $a=2$ in Fig. 11).
After that, from $-10$ to $2$ dB, all the power will be allocated to
the direct link $S-D$ and without using the two-hop transmission
again. Hence, there will be no opportunity to use the relay to
forward the information. As a result, there will be no opportunity
to use the signal time coding in the time domain. After 3 dB, all
the transmission power will be allocated to the direct link $S-D$
and $S-R$ again. Hence, there will have an opportunity to use the
relay to forward the information again.  As a result, there will
have an opportunity to use the signal time coding in the time domain
and the information flow rate gain by using signal time coding in
relatively high SNR will become greater.

Observe the case $a=1.5$, when SNR is very low, all the power will
be allocated to link $S-R$ and not to the direct link of $S-D$. This
is similar to that of $a=2$. When SNR is higher than -9 dB, the
optimal power allocation will be balanced between the link $S-R$ and
$S-D$.  This indicates that there will have an opportunity to use
the relay to forward the information. As a result, there will have
an opportunity to use the signal time coding in the time domain to
get a higher information rate.

 The results in Fig. 11 also show that
when SNR is very high, the optimal power allocation will have the
same trends which mainly be dependent on the relative distance ratio
$\kappa$.

We conclude this section by noting  that in the three node network
model, the mixed transmission by using two-hop and direct
transmissions is not always a good option. That is, only two-hop
transmission, only direct link transmission and the mixed
transmission may occur, which is dependent on the practical
scenarios.

\section{Conclusion}
In this paper, we first derived  explicit formulas to accurately
estimate the maximum number of available codes by using an iterative
algorithm for an arbitrary code length of signal-time coding, and
then presented an iterative construction method of codebooks.
Compared to exhaustive computer search, the proposed method can
reduce the complexity greatly.  In addition and to avoid having the
codebook storing complexity increasing exponentially  with
code-length as well as reduce the buffer size while keeping a
relatively high efficiency, we proposed a combinatorial universal
construction method. Furthermore, we considered the application of
signal-time coding in wireless Gaussian relay networks. It was shown
that it can get a higher information rate than the upper bound using
conventional signal transmission technique in some scenarios. An
optimal power allocation analysis was presented and it was found
that in the three node network model, the mixed transmission by
using two-hop and direct transmissions is not always a good option

\section*{Acknowledgments}
This work was partially supported by NSFC/RGC joint grant No.
60831160524.

%\begin{figure}[h!]
%\centering
%\includegraphics[width = 10.0cm]{timeflow.eps}
%\caption{Information Transmission flows by using signal-time coding
%}\label{fig_2}
%\end{figure}

\begin{table}[h!]
\centering
\caption{Codebook construction with the highest coding
efficiency }\label{table}
\includegraphics[width = 20.0cm]{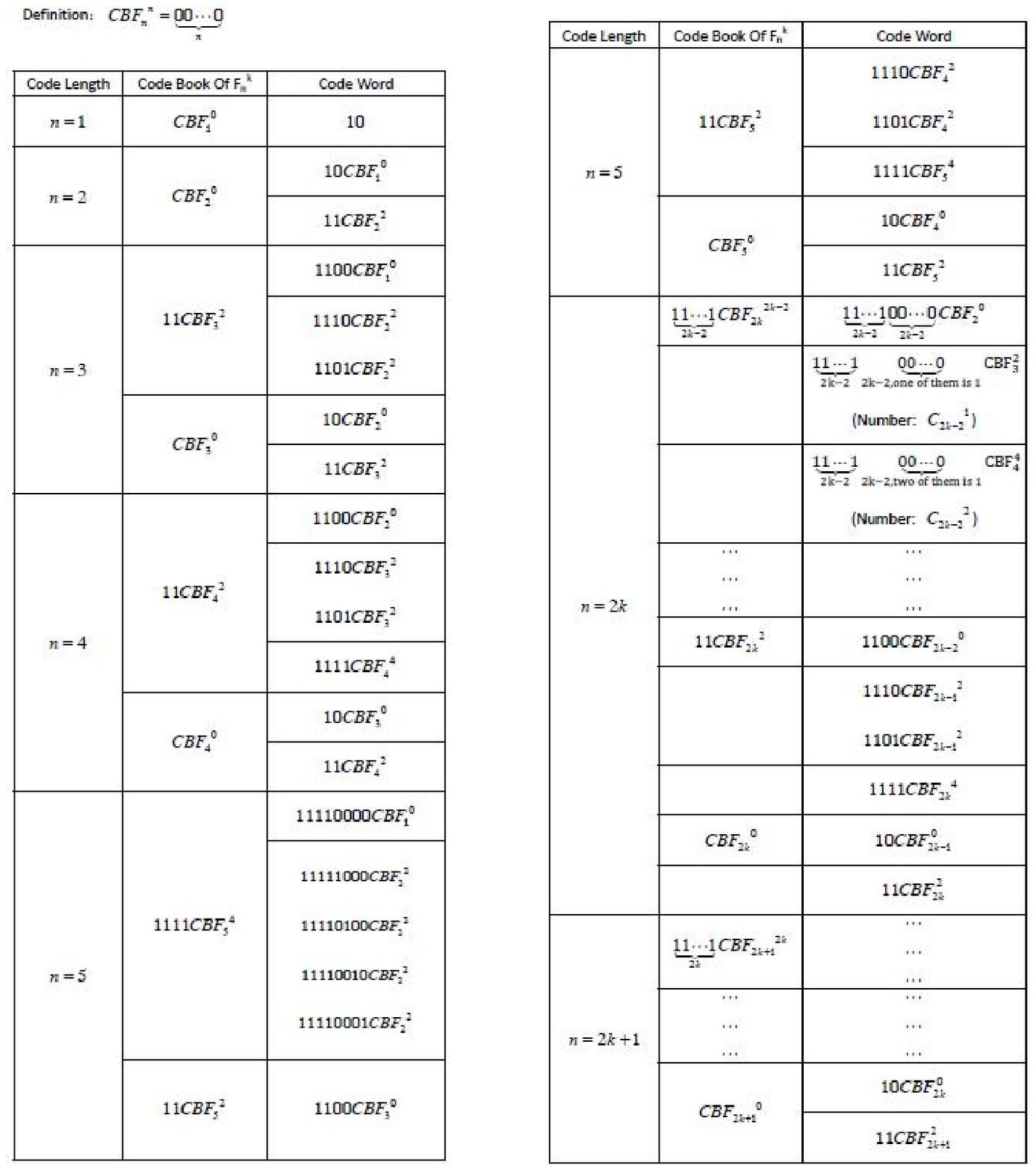}
\end{table}
\begin{figure}[h!]
\centering
\includegraphics[width = 10.0cm]{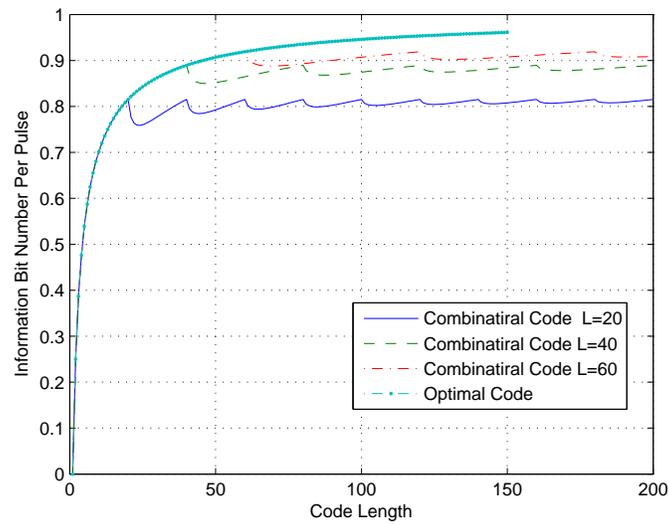}
\caption{Coding rate per pulse for different codebooks
}\label{fig_4}
\end{figure}

\begin{figure}[h!]
\centering
\includegraphics[width = 10.0cm]{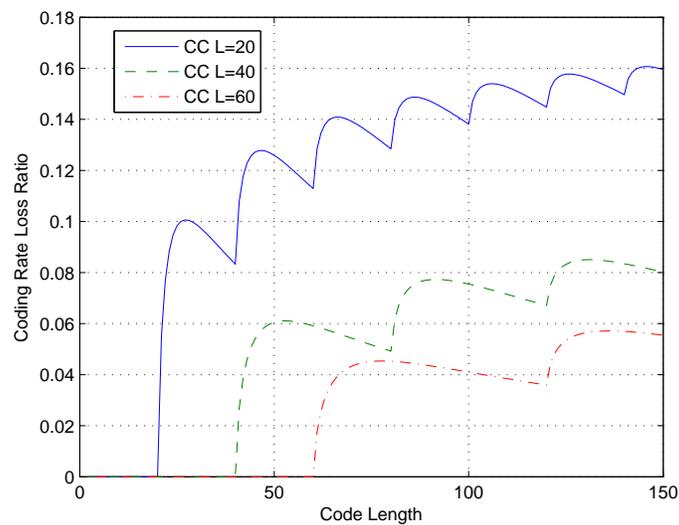}
\caption{Coding rate loss ratio of Concatenated codes}\label{fig_5}
\end{figure}
\begin{figure}[h!]
\centering
\includegraphics[width = 10.0cm]{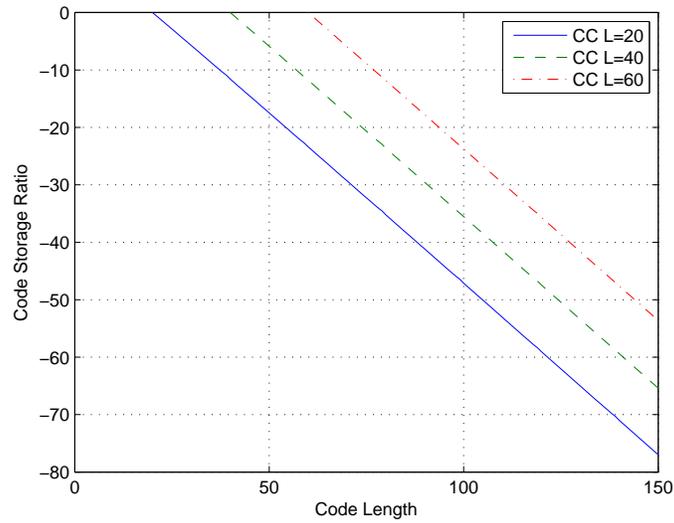}
\caption{Code storage ratio of Concatenated codes}\label{fig_6}
\end{figure}

\begin{figure}[h!]
\centering
\includegraphics[width = 10.0cm]{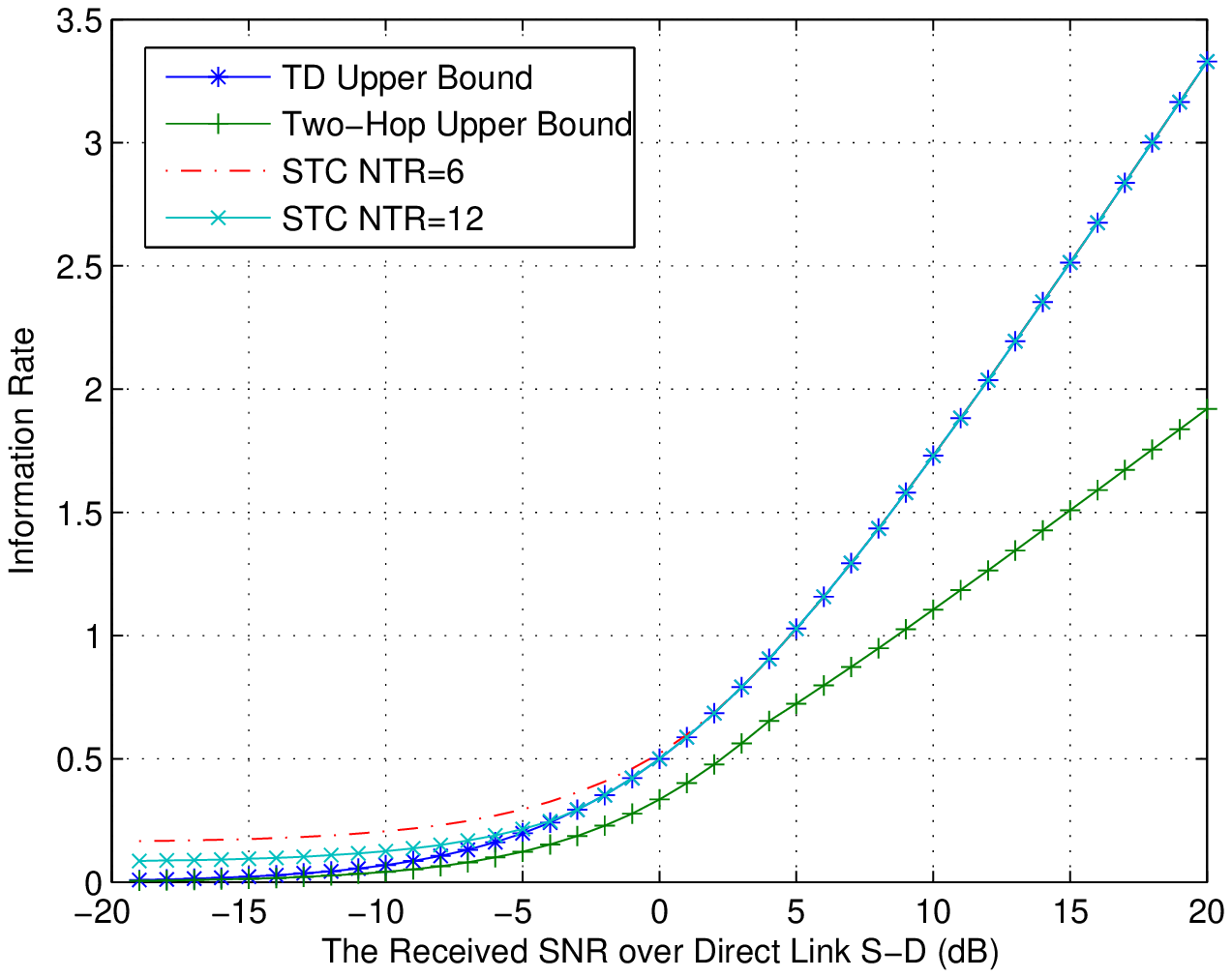}
\caption{Information Rate comparison, where $\kappa=0.35$, $a=2$ and
the normalized time resolution (NTR) $B\Delta T =6$ and
12}\label{fig_7}
\end{figure}

\begin{figure}[h!]
\centering
\includegraphics[width = 10.0cm]{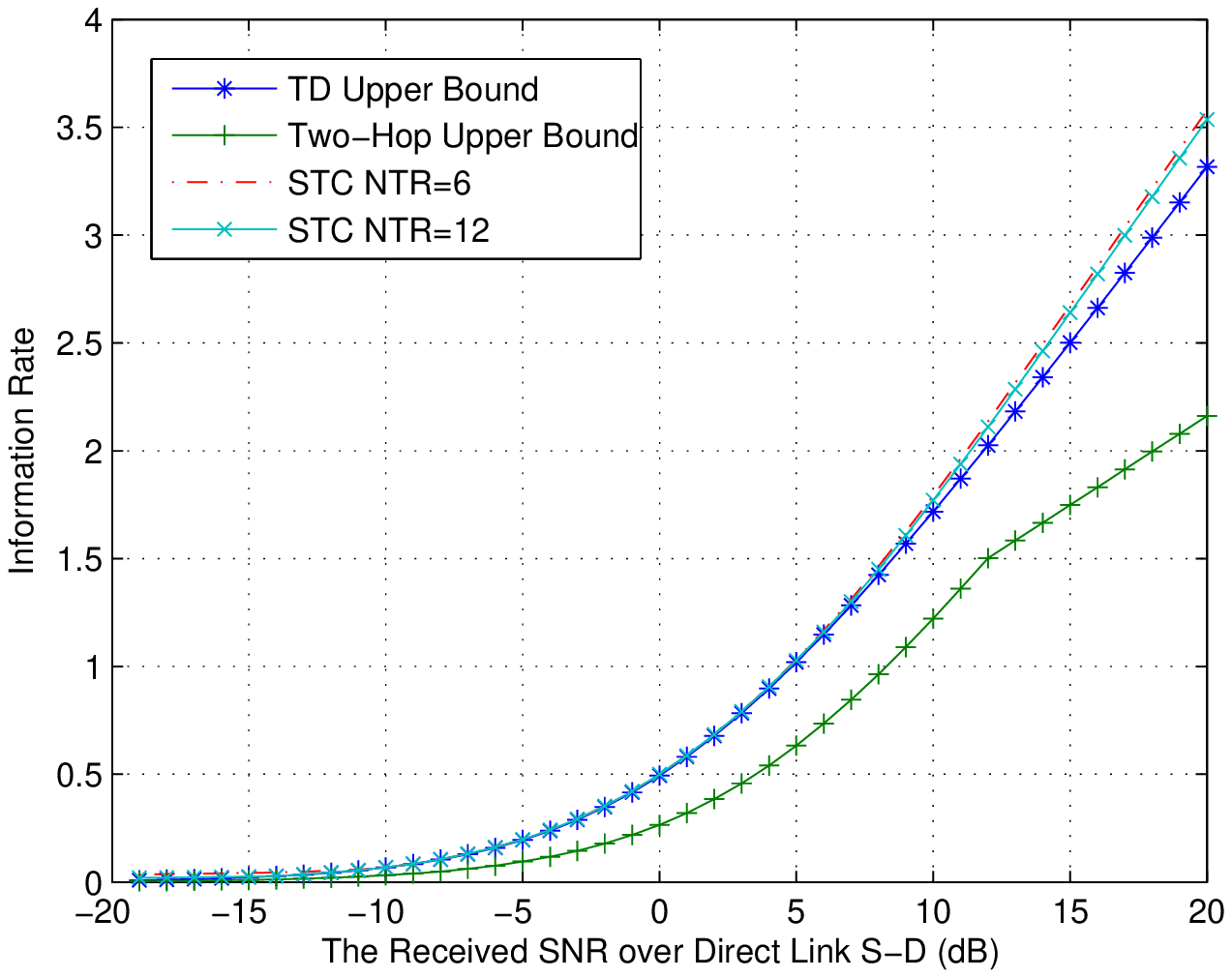}
\caption{Information Rate comparison, where $\kappa=0.75$, $a=2$
and the normalized time resolution (NTR) $B\Delta T =6$ and
12}\label{fig_7}
\end{figure}

\begin{figure}[h!]
\centering
\includegraphics[width = 10.0cm]{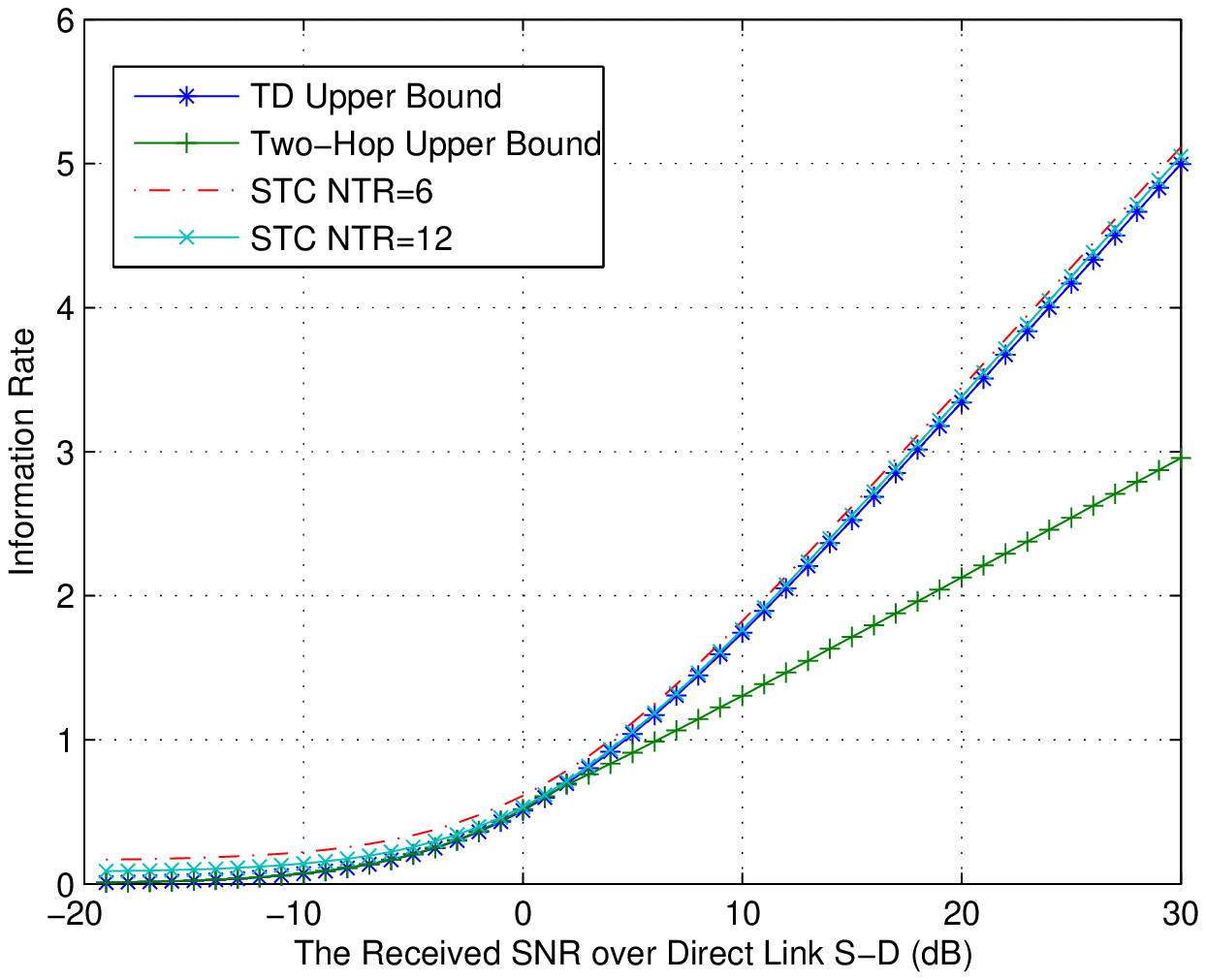}
\caption{Information Rate comparison, where $\kappa=0.35$, $a=1.5$
and the normalized time resolution (NTR) $B\Delta T =6$ and 12}
\label{fig_9}
\end{figure}

\begin{figure}[h!]
\centering
\includegraphics[width = 10.0cm]{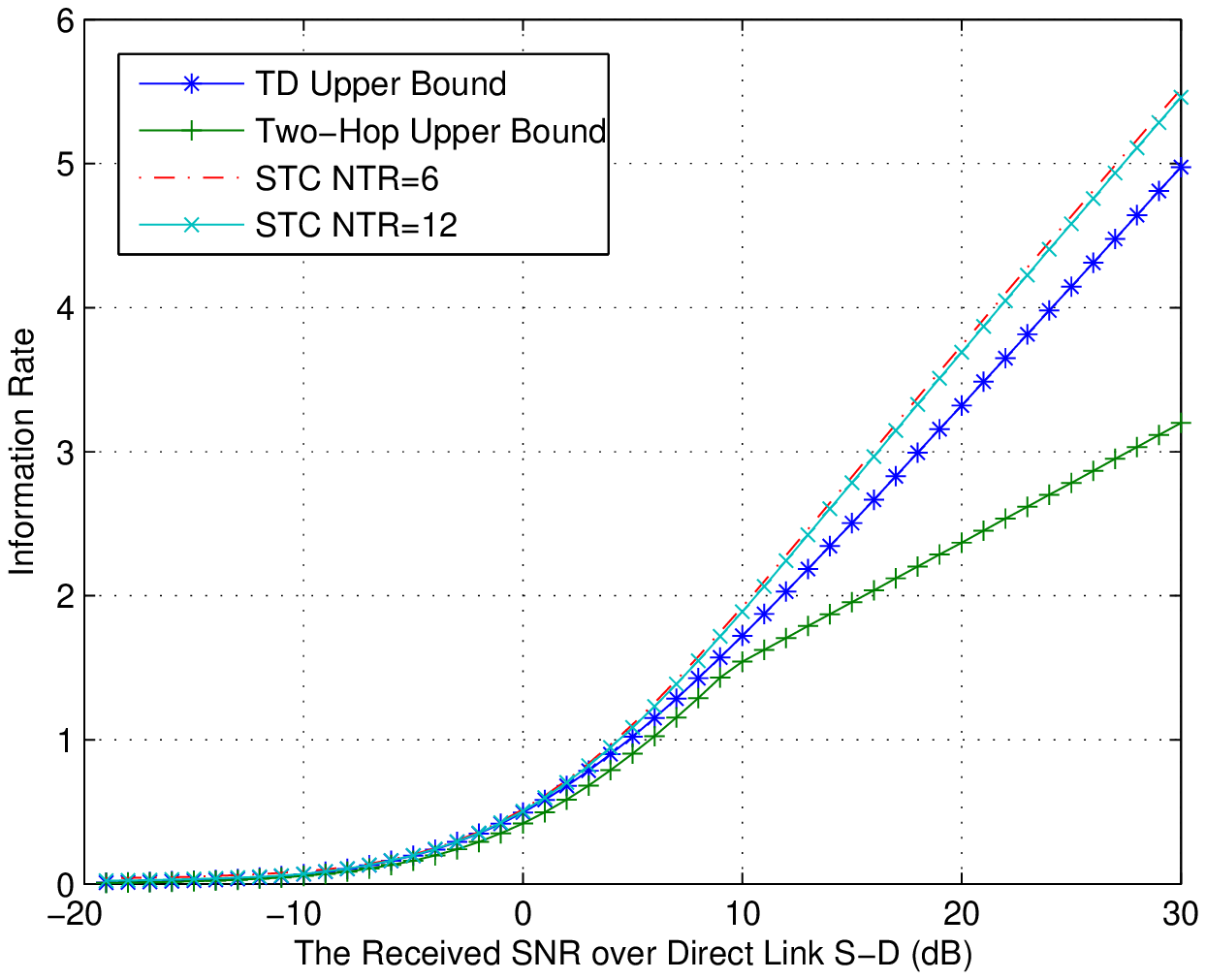}
\caption{Information Rate comparison, where $\kappa=0.75$, $a=1.5$
and the normalized time resolution (NTR) $B\Delta T =6$ and 12}
\label{fig_10}
\end{figure}

\begin{figure}[h!]
\centering
\includegraphics[width = 10.0cm]{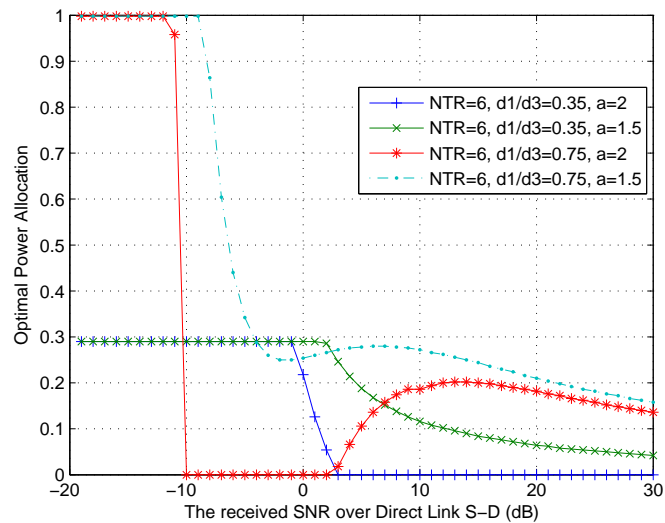}
\caption{Optimal power allocation in different scenarios}
\label{fig_11}
\end{figure}
\end{document}